# A Novel Uncertainty Parameter SR (Signal To Residual Spectrum Ratio) Evaluation Approach For Speech Enhancement

M. Ravichandra Kumar[1] and B. Ravi Teja[2]

[1]Department of Electronics and Communication,
M-tech, Gudlavalleru Engineering College, A.P, India
[2]Department of Electronics and Communication Engineering, Assistant professor,
Gudlavalleru Engineering College, A.P, India

## ABSTRACT

*Usually, hearing impaired people use hearing aids which are implemented with speech enhancement algorithms. Estimation of speech and estimation of nose are the components in single channel speech enhancement system. The main objective of any speech enhancement algorithm is estimation of noise power spectrum for non stationary environment. VAD (Voice Activity Detector) is used to identify speech pauses and during these pauses only estimation of noise. MMSE (Minimum Mean Square Error) speech enhancement algorithm did not enhance the intelligibility, quality and listener fatigues are the perceptual aspects of speech. Novel evaluation approach SR (Signal to Residual spectrum ratio) based on uncertainty parameter introduced for the benefits of hearing impaired people in non stationary environments to control distortions. By estimation and updating of noise based on division of original pure signal into three parts such as pure speech, quasi speech and non speech frames based on multiple threshold conditions. Different values of SR and LLR demonstrate the amount of attenuation and amplification distortions. The proposed method will compared with any one method WAT(Weighted Average Technique) Hence by using parameters SR (signal to residual spectrum ratio) and LLR (log like hood ratio), MMSE (Minim Mean Square Error) in terms of segmented SNR and LLR.*

## KEYWORDS

*Noise Estimation, Voice Activity Detector (VAD), Speech Enhancement, SR (Signal to Residual spectrum ratio) parameter, Speech Intelligibility Improvement.*

## 1. INTRODUCTION

The major problem arises in speech enhancement background noise and it is affected by speech signal. There are many applications which are speech recognition, hearing aid, VOIP (Voice over Internet Protocol), teleconferencing systems and mobile phones of reduces background noise. The noise present in the both analogy and digital systems. An unwanted signal as noise and it degrades the speech intelligibility and speech quality. Vehicle noise and background noise are the different types of noises. In speech enhancement mainly considered as noise estimation it requires to estimate of noise from noisy speech signal. Speech enhancement main objective is to give better performance of speech quality and speech intelligibility by using various algorithms and based on these algorithms to minimise the MSE (Man Square Error) [5]. The effect of various distortions (attenuation and amplification distortions) present in the speech signal so these distortions are proper control to improve the speech intelligibility. The negative difference between clean and enhanced spectrum would be amplification distortion, while a positive difference would be attenuation distortion. Speech enhancement for noise reduction can be categorised into three fundamental classes and those are model based, spectral restoration and filtering technique methods. All the methods are common feature is clean speech power spectrum estimation from noisy environment spectrum.

The presence or absence of human speech detected is called VAD (Voice Activity Detector). In speech processing technique used VAD and also called as speech detection or speech active detector as well as VAD used in noise reduction also. Multimedia application VAD allows simultaneously voice and data. Here consider another application cellular based system (GSM, CDMA) in discontinuous transmission mode. Speech intelligibility and speech quality both are correlated highly by measuring frequency domain of segmental SNR so for this measure is refer to residual spectrum ratio [14].

## 2. RELATED WORK

In obtainable algorithms are not suitable for estimate of background noise but VAD (voice activity detector) good background noise estimation algorithm for stationary environment [13]. Speech presence or speech absence of human speech is detected by VAD (voice activity detector) by using this algorithm to estimate noise in speech pauses only. Every algorithm makes to give speech quality but not speech intelligibility and this drawback occurred in present existing algorithms [3]. Wiener and MMSE (minimum mean square error) algorithms are used to minimize the error in between of enhanced and clean spectrum so these algorithms are based in spectral principals.

Most of the algorithms were proposed speech recognized application to estimate the noise in non stationary environments VAD did not estimate the noise accurately. The lack of intelligibility in present algorithms is not proper to estimation of noise. These problems can be reduces by using the propose algorithm SR (signal to residual spectrum ratio) for improve speech quality and speech intelligibility in noisy environment.

## 3. PROPOSED WORK

Consider, here P(n) and Q(n) are the clean speech, noise and then noisy speech denoted as follows,

$$X(n) = P(n) + Q(n) \tag{1}$$

Time domain of noisy speech is segmented by frames by using of windowed technique let it be consider hamming window and represented equation as follows

$$W[n,\tau] = 0.54 - 0.4 \cos\left(\frac{2\Pi(n-1)}{N_w-1}\right) \quad \text{for } 0 \leq n \leq N_{w-1} \tag{2}$$

The short time Fourier transforms is give equation for wave form of windowed speech signal

$$S[\omega,\tau] = \sum_{n=\infty}^{\infty} ,\tau] \exp[-j\omega n] \tag{3}$$

Where τ represents centre time at window

### 3.1 Determination of Threshold Condition

Speech intelligibility and speech quality both are correlated highly so to measure using segmental SNR (Signal to Noise Ratio) in consider frequency domain version and this measure to mention as signal to residual spectrum.

$$SNR_{ESI(k)} = \frac{S^2(k)}{(S(k)-\hat{S}(k))^2} \tag{4}$$

$\hat{S}(k)$ is speech enhancement algorithm of estimated spectrum and $S(k)$ is clean speech magnitude spectrum. To improve the speech intelligibility by proper control of distortions using regions are constraint and it has follows

a) $S(k) \leq \hat{S}(k)$, suggested only attenuation distortion
b) $S(k) \geq 2. \hat{S}(k)$, suggested greater or 6.02 db of amplification distortion
c) $S(k) \leq \hat{S}(k) \leq 2. S(k)$ , suggested up to 6.02 db of amplification distortion

Reason (a) and Reason (b) from that we constraint to this reason $\hat{S}(k) \geq S(k)$ and it is used in speech enhancement algorithms.

$$\hat{S}(k) \geq 2. S(k) \tag{5}$$

This after squaring on both sides becomes

$$\hat{S}^2(k) \geq 4. S^2(k) \tag{6}$$

So assume $\hat{S}(k)= X(k)$ it is not enhance noisy speech by algorithms and then $\hat{S}^2(k) = X^2(k) = \hat{S}^2(k) + Q^2(k)$ and reduces to $\hat{S}^2(k) \geq 1/3 \ Q^2(k)$

$$SNR(k) \geq 1/3 \tag{7}$$

### 3.2. SR (Signal to Residual spectrum ratio)

Figure.1 represents the SR algorithm and the noisy signal is segmented using windowed technique eq.(2) later FFT is performed on the segmented frames with the help of with the help of eq.(3). Noisy speech has different frames so we can calculate SNR (Signal To Noise Ratio) based on threshold determination.

### 3.3. Noise power estimation method

Here focused on noise estimation and it has different approaches so the fundamental component of speech enhancement is noise power estimation. It required estimating of noise from noisy speech spectrum by using different algorithms based on classification of speech into quasi speech, original speech and noise speech [11].

### 3.3.1. Non-Speech

It has to be occurred in speech absence or speech pauses only and to estimate noise power of these frames for the following proposed condition

if $\hat{S}(k) \geq 2.S(k)$ then

$$A(m,k) = \alpha \ \hat{A}(m-1,k)+(1-\alpha) \ |\hat{A}(m,k)|^2 \tag{8}$$

Where α is called as smoothing factor and typically set to α=0.98 and lies in 0<α<1.

### 3.3.2. Quasi-Speech

To estimate noise power for quasi speech is both noise and speech on each frame and the proposed condition

It $S(k) \leq \hat{S}(k) \leq 2. S(k)$ then
$$\hat{A}(m, k) = B(m, k)\hat{A}(m-1, k)+(1-B(m, k)) \qquad (9)$$

Where $\hat{A}(m, k)$ is non speech frame of noise spectrum estimation

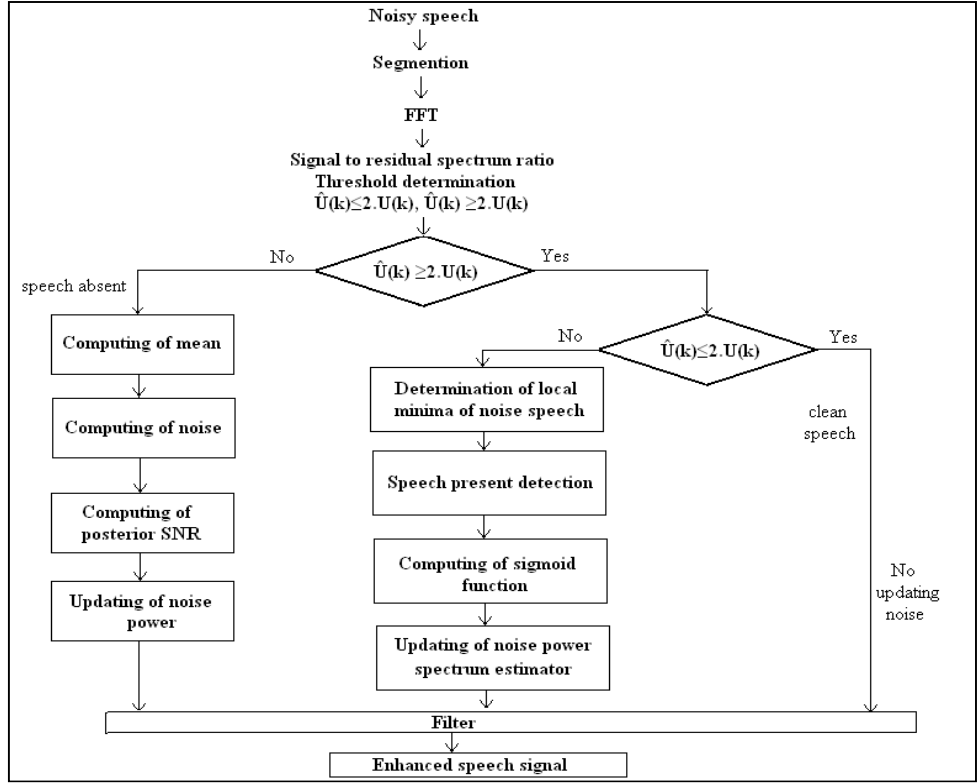

Figure 1: SR Algorithm.

## 3.4. Tracking the Minimum of Noisy Speech

To tracking of noisy speech by regularly averaging precedent spectral values, here used rule non linear in different approach [10]

if $B_{min}(m-1, k) \leq B(m, k)$ then
$$B_{min}(m, k) = \gamma B_{min}(m-1, k) + \frac{1-\gamma}{1-\beta}(B(m, k) - \beta B(m-1, k)) \qquad (10)$$

If $B_{min}(m-1, k) > B(m, k)$ then
$$B_{min}(m, k) = B(m, k) \qquad (11)$$

To determine the values of γ, β and ξ by experiment, in practical implementation smoothing parameter in (11) whose maximum value is 0.98 to avoid deadlock for $r(m,k)=1$.

## 3.5. Speech Presence Probability

To measure how much speech present probability in noisy speech by following equation

$$B_{sp}(m,k) = \frac{|A(m,k)|^2}{B_{min}(m,k)} \qquad (12)$$

where $B_{min}(m,k)$ and $|A(m,k)|^2$ are represented as local minimum and power spectrum of noisy speech. Speech present and speech absent are dependent on the ratio of speech present probability if it is grater to threshold then consider as speech present otherwise it gives speech absent.

### 3.6. Computing Logistic Function

Logistic function is one of special case in the form of mathematical and it is also called as sigmoid function or sigmoid curve as given function

$$g(x)=1/(1-e^{-x}) \qquad (13)$$

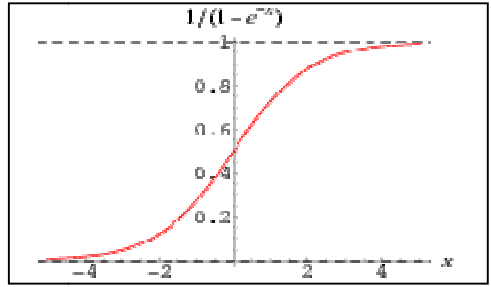

Figure 2: sigmoid curve

### 3.7. Calculating Frequency Dependent Smoothing Constant

To compute smoothing factor need the time-frequency domain as to follow this equation [9].

$$B(m, k) = \alpha_s + (1-\alpha_s)\, Bsp(m, k) \qquad (14)$$

Where, $\alpha_s$ is denotes as constant.

To updating of minimum noisy spectrum is $B_{min}(m, k)$ and given equation

$$B(m, k) = \xi B(m-1, k) + (1-\xi) \qquad (15)$$

Where, $B(l, k)$ is average noise spectrum and $\xi$ is known as smoothing factor

$$b(\lambda, k) = \alpha_b\, b(\lambda-1,k)+(1-\alpha_b)\, I(\lambda, k) \qquad (16)$$

Where $b(\lambda, k)$ is a smoothing constant, the above recursive absolutely utilize the correlation for speech presence in adjacent frames.

For $r(m, k) = 1$.

$$r(m, k) = N(m-1, k)/\sigma_N^2(m, k) \qquad (17)$$

Posterior SNR of smoothed version is represented by eq. (16)

The Wiener filter solves the signal estimation problem for stationary signals. A major contribution was the use of a statistical model for the estimated signal the filter is best in the intellect of the MMSE [16]. We shall focus here on the discrete-time version of the Wiener filter and it is used to generate estimated pure signal from a given noise speech signal.

## 4. IMPLEMENTATION AND RESULTS

Speech enhancement algorithms are tested on MATLAB for Non-stationary environment of speech database [2]. The unvoiced speech regions to detected correctly by observed the results of proposed algorithm and speech activity region also accurately measured even noise is present. The table gives classification of results and performance of the algorithms in which segmental SNR and LLR (Log Like hood Ratio) are compared to proposed algorithm SR (Signal to Residual spectrum ratio). Spectrogram is way to visualize the speech signal in the domain time-frequency representation. In speech signal through several intermediate levels which are linguistic message and paralinguistic information including emotion is effectively visualized based on the spectrogram. Now we can concludes the variations in the noisy speech signal of Spectrogram represented in different areas those are trains, cars, and airport.

Table1: comparison of weighted average technique and proposed SR technique using LLR and segmental SNR methods

| Type of Noise | SNR in db | LLR | | Segmental SNR | |
|---|---|---|---|---|---|
| | | Weighted Average Technique | Proposed (SR) Technique | Weighted Average Technique | Proposed (SR) Technique |
| CAR | 0 | 1.687827 | 1.500914 | -6.806391 | -6.716270 |
| | 5 | 1.842711 | 1.596159 | -5.668619 | 5.485975 |
| | 10 | 1.976017 | 1.602708 | -4.866237 | -3.861581 |
| | 15 | 1.831509 | 1.580956 | -4.335797 | -3.537122 |
| AIRPORT | 0 | 1.237398 | 1.057377 | -3.802483 | -3.440414 |
| | 5 | 1.124488 | 0.934859 | -2.781458 | -2.526855 |
| | 10 | 0.919158 | 0.736983 | -0.731036 | -0.083965 |
| | 15 | 0.910468 | 0.549913 | 1.310788 | 3.080826 |
| TRAIN | 0 | 2.091845 | 1.798190 | -6.486321 | -6.296185 |
| | 5 | 2.322675 | 1.845213 | -5.559169 | -4.970945 |
| | 10 | 2.036162 | 1.759774 | -5.251629 | -4.206358 |
| | 15 | 2.230337 | 1.827800 | -3.211548 | 4.284449 |

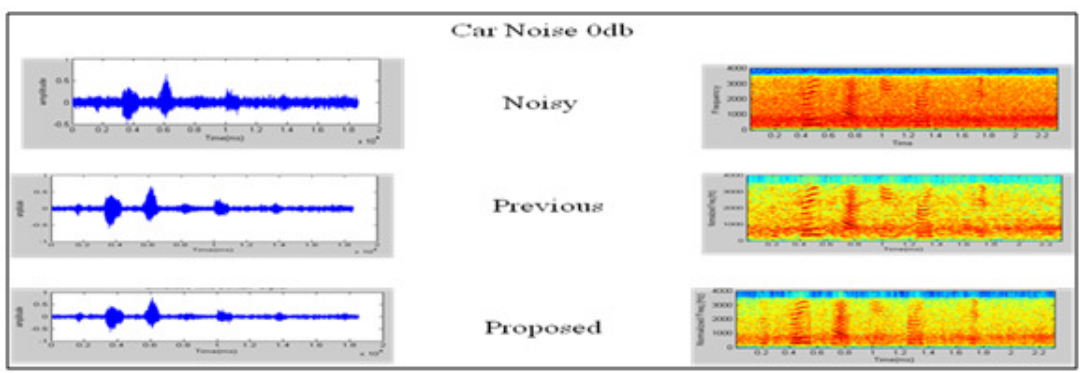

(i)

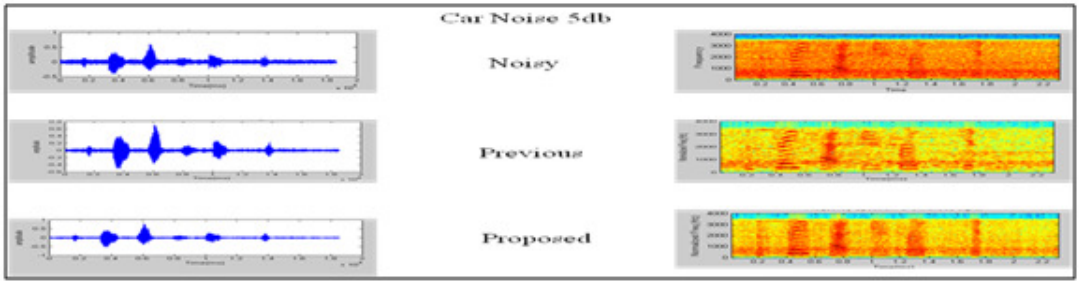

(ii)

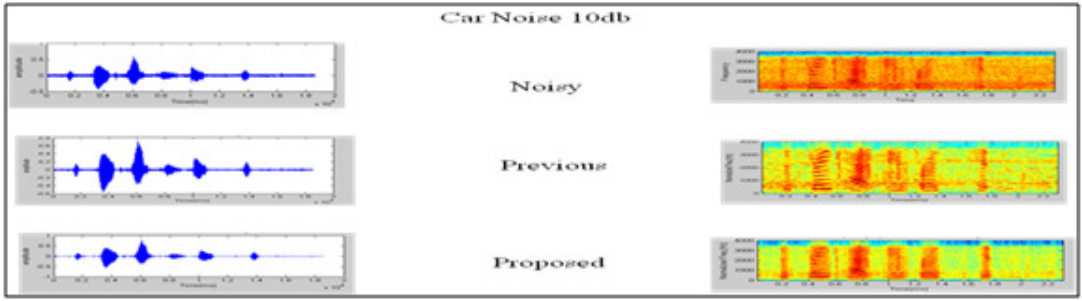

(iii)

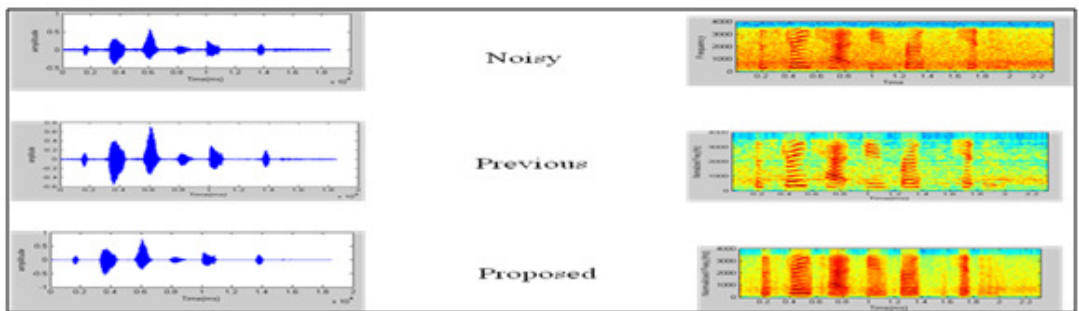

(iv)

Figure 3: timing wave form and spectrogram of (i) pure speech signal (ii) noisy speech signal and enhanced signal with (iii) weighted average technique (iv) proposed SR technique in car noise with different SNR levels.

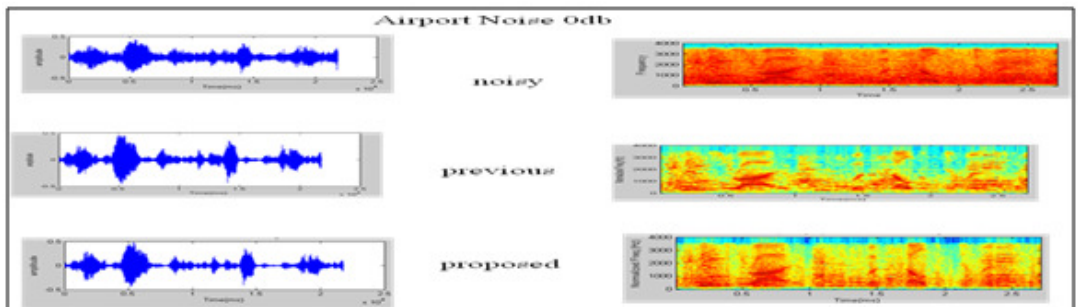

(i)

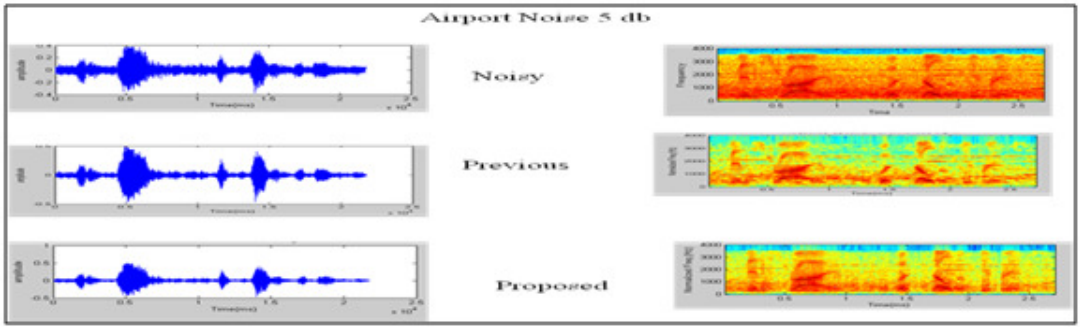

(ii)

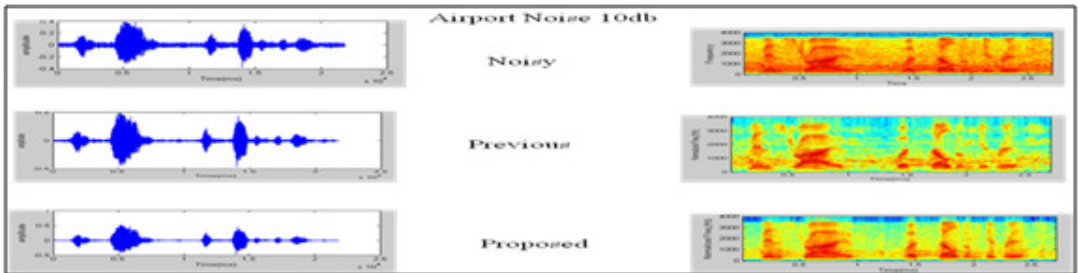

(iii)

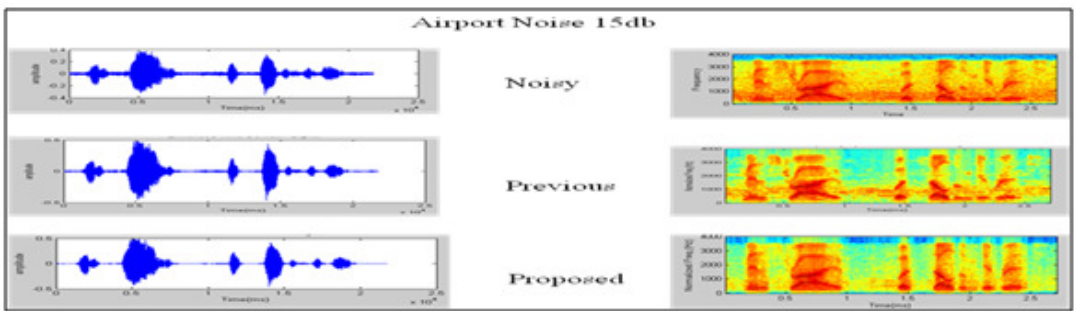

(iv)

Figure 4: timing wave form and spectrogram of (i) pure speech signal (ii) noisy speech signal and enhanced signal with (iii) weighted average technique (iv) proposed SR technique in airport noise with different SNR levels**.**

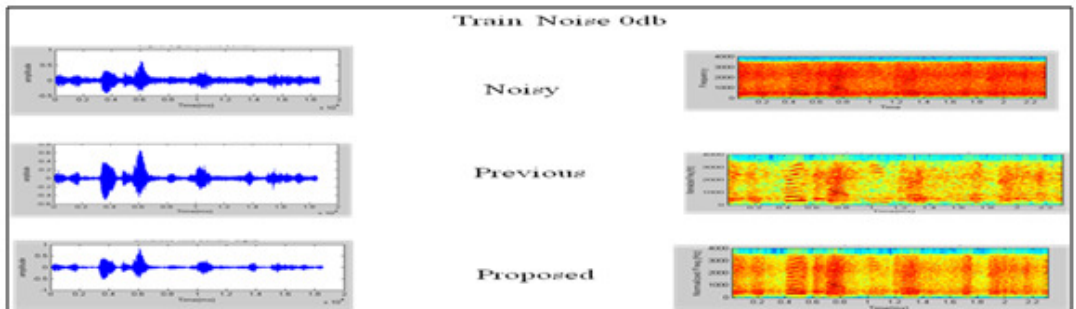

(i)

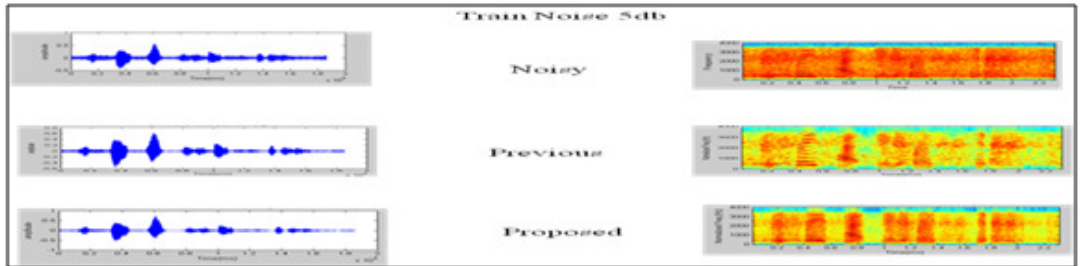

(ii)

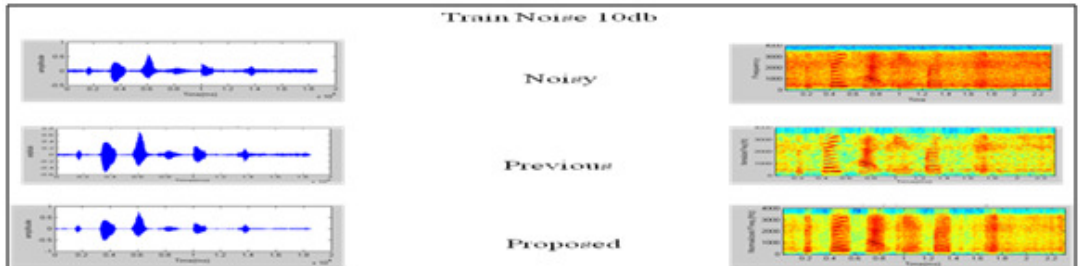

(iii)

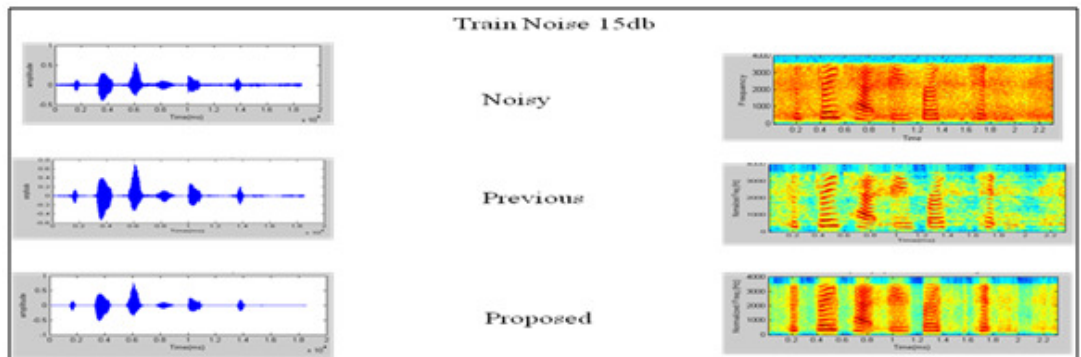

(iv)

Figure 5: Timing wave form and spectrogram of (i) pure speech signal (ii) noisy speech signal and enhanced signal with (iii) weighted average technique (iv) proposed SR technique in Train noise with different SNR levels.

## 5. CONCLUSION

This paper focused on the issue of noise estimation for enhancement of noisy speech. The noise estimate was updated continuously in every frame using time–frequency smoothing factors calculated based on speech-presence probability in each frequency bin of the noisy speech spectrum [1].The main achievements of speech enhancement algorithms are speech intelligibility and speech quality. Here to reduce the amplification distortion and attenuation distortion by using proposed method SR (Signal to Residual spectrum ratio) [5]. The proper control of these distortions to improve speech intelligibility it is main drawback of speech enhancement algorithms. The proposed method SR it gives better performance when compared to the previous existing methods are LLR (log like hood ratio) and segmental SNR.

## Authors

**Ravichandra Kumar Manike** pursuing M.Tech in the branch of Digital Electronics and Communication Systems at Gudlavalleru Engineering College and B.Tech degree in Electronics and Communication Engineering received from Prakasam Engineering College in the year of 2011.Gate qualified in the year 2012 &13.

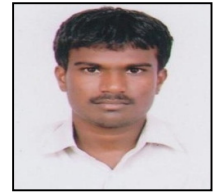

**Ravi Teja Ballikura** received the B.Tech and M.Tech degree in Electronics and Communication Engineering in 2010 from Bapatla Engineering College, Digital electronics and communication systems in 2012 from Gudlavalleru Engineering College affiliated by JNTUK, Kakinada respectively. Working as a assistant professor in Gudlavalleru Engineering College from 2012 to till date. Research interests in speech processing and more especially in enhancement of speech signal.

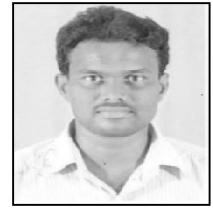